\documentclass[sigconf]{acmart}

\newenvironment{rachael}{\color{black}}{}

\definecolor{blue}{named}{black}

\AtBeginDocument{%
  }

\copyrightyear{2026}
\acmYear{2026}
\setcopyright{cc}
\setcctype{by}
\acmConference[CHI '26]{Proceedings of the 2026 CHI Conference on Human Factors in Computing Systems}{April 13--17, 2026}{Barcelona, Spain}
\acmBooktitle{Proceedings of the 2026 CHI Conference on Human Factors in Computing Systems (CHI '26), April 13--17, 2026, Barcelona, Spain}
\acmPrice{}
\acmDOI{10.1145/3772318.3791658}
\acmISBN{979-8-4007-2278-3/2026/04}

\begin{document}

\title{Hiding in Plain Sight: Understanding the Everyday Practices and Challenges of Car Dwellers}

\author{Rachael Zehrung}
\orcid{0000-0003-1617-9079}
\affiliation{%
  \institution{Department of Informatics \\ University of California, Irvine}
  \city{Irvine}
  \state{CA}
  \country{USA}
}
\email{rzehrung@uci.edu}

\author{Yunan Chen}
\orcid{0000-0003-4056-3820}
\affiliation{%
  \institution{Department of Informatics \\ University of California, Irvine}
  \city{Irvine}
  \state{CA}
  \country{USA}
}
\email{yunanc@ics.uci.edu}


\begin{abstract}
Vehicle dwelling has increased significantly in recent years. While HCI research has explored vehicle dwelling through the lens of digital nomadism and vanlife, it has largely overlooked the complexities of vehicle dwelling as a form of housing insecurity, as well as the unique constraints of living in smaller vehicles. Drawing on a qualitative analysis of posts and comments from an online community, we examine car dwellers' infrastructuring work to manage daily life under social, spatial, and infrastructural constraints. We further explore the motivations and identity negotiations of car dwellers, whose experiences fall between homelessness and nomadism, and highlight how developing infrastructural competence can shape identity. We discuss implications for future HCI research on mobility and dwelling under conditions of uneven access to infrastructure and provide design recommendations for technologies that better account for car dwellers' diverse needs, circumstances, and identities.



\end{abstract}

\ccsdesc[500]{Human-centered computing~Empirical studies in HCI}
\ccsdesc[500]{Human-centered computing~Empirical studies in collaborative and social computing}

\keywords{Vehicle Dwelling, Homelessness, Mobility, Infrastructure}

\maketitle

\section{Introduction}
Vehicle dwellers describe individuals who live in cars, vans, and recreational vehicles as their primary or part-time residence. Driven by rising housing costs and the expansion of remote work, vehicle dwelling has increased substantially in the United States since the pandemic, with an estimated 3 million individuals living in vans in 2022~\cite{Statista2022VanLifers}. This phenomenon exists at the intersection of two distinct narratives: social media's romanticization of \#vanlife as an adventurous nomadic lifestyle~\cite{rizvi2021provocations,gretzel2019vanlife}, and growing public concern about vehicle dwelling as a form of homelessness that increases individuals' risk of poor physical and mental health outcomes, victimization, and stigmatization~\cite{giamarino2024lives,pruss2022long,richards2023unsheltered}.

Information and communication technologies have demonstrated potential for supporting people experiencing homelessness (PEH) by helping them access information and services, maintain social relationships, and manage their identity~\cite{woelfer2012homeless,burrows2019evaluating,le2008designs}. However, most of this work has centered on individuals living on the streets or in shelters, with few studies focused specifically on vehicle dwellers as a distinct subgroup. In a separate body of work, HCI research has investigated vehicle dwelling through the lens of digital nomadism and vanlife culture, highlighting nomads' extensive use of technology and the infrastructuring work they perform to manage everyday life~\cite{rizvi2021provocations,lee2019social,sutherland2017gig}. Both PEH and digital nomads experience disruptions to familiar infrastructures and routines, and technology can support the process of adapting and learning how to manage daily life again. 

These parallel bodies of work have yet to converge around the unique needs of vehicle dwellers, who occupy a position along the spectrum between nomadism and homelessness. Little is known about the specific challenges, practices, and needs of this population, particularly individuals living in vehicles out of necessity rather than choice. Studying vehicle dwelling is complicated by its nature as a form of ``hidden'' homelessness, meaning that vehicle dwellers often remain invisible to service providers and undercounted in official estimates~\cite{deleu2023hidden,crawley2013needs}. Individuals living in cars (e.g., sedans) are even less visible due to the smaller size of their vehicles and ability to blend into urban environments. Cars also present distinct challenges as living spaces due to spatial constraints and limited conversion possibilities. These factors make car dwelling a critical but overlooked site of inquiry for understanding the realities of vehicle dwelling, especially as a form of housing insecurity. \textcolor{blue}{We guided this inquiry with the following research questions: }

\begin{itemize}
    \item \textcolor{blue}{How do car dwellers manage everyday life? What challenges do they face and what strategies do they use? What role does technology play in shaping car dwellers' practices and routines?}
    \item \textcolor{blue}{What are car dwellers' motivations for car living, and how do they make sense of their experiences and identities?}
\end{itemize}

To address \textcolor{blue}{these questions}, we investigate the lived experiences and perspectives of car dwellers through a thematic analysis of posts and comments on r/urbancarliving, one of the largest and most active online communities for car dwellers. Our analysis uncovered rich insights about users' everyday challenges with car living and the infrastructuring work they performed to manage daily life. Our work makes the following contributions to HCI: 

\begin{itemize}

   \item An empirical understanding of car dwellers' practices to manage daily life inside and outside the vehicle, as well as the role of different technologies in supporting their infrastructuring work.

    \item A conceptual understanding of car dwelling as a life in between, neither fully aligned with homelessness nor with nomadism. We identified two primary motivations for car living, distinguishing between users who turned to car dwelling as a temporary necessity and those who embraced it as a lifestyle choice. While initial motivations for car living shaped users' perspectives of their living situation, we highlight how developing infrastructural competence can shift identity.
   
    \item \textcolor{blue}{Insight into individuals' infrastructuring work to resolve breakdowns across multiple infrastructures, particularly in light of uneven access to resources and public infrastructures. Building on this understanding, we discuss implications for future studies on mobility and dwelling.}

    \item \textcolor{blue}{Design recommendations for technologies that better support the diverse needs and identities of car dwellers, with a focus on individuals living in their cars out of necessity or who identify as homeless.}

\end{itemize}

\section{Related Work}

\subsection{Homelessness and Technology}
Prior work has found that information and communication technologies (ICTs) contribute to the overall wellbeing of people experiencing homelessness (PEH) by connecting them to essential services, fostering social connectedness and inclusion, and supporting a sense of safety~\cite{le2008designs,roberson2010survival,woelfer2011improving}. Through mobile applications and the internet, PEH can navigate the demands of homelessness by finding and securing basic needs such as food, employment, and housing~\cite{burrows2019evaluating,chandra2021critical,mohan2019food}. In addition, PEH often seek social support online, particularly around stigmatized topics such as mental health conditions, substance use disorders, and sexual health~\cite{curry2016correlates,barman2011sexual}, which intersect with the stigmatization of homelessness~\cite{thomas2022negotiating}. Social media can also help PEH maintain and cultivate pro-social ties, manage their identities, and receive emotional and practical support~\cite{hu2019characterizing,le2008designs,woelfer2012homeless}. At the same time, PEH face barriers to technology use, including inconsistent access to data plans and the internet, limited ability to charge devices, and a shortage of private, secure places to use and store personal devices~\cite{woelfer2011homeless,sleeper2019tough}. Together, these studies highlight both the value of ICTs for supporting PEH and the infrastructural constraints that shape their use. While this body of work offers valuable insights, it primarily focuses on individuals living in shelters or outdoors. Car dwellers occupy a different context, as their vehicles provide a degree of privacy and infrastructure not available in other unsheltered situations while also introducing unique challenges. Our work extends HCI research on homelessness by examining the technology practices and needs of car dwellers, an overlooked and undersupported subgroup of PEH.

\subsection{Mobile Living and Vanlife}
HCI research has long examined the technology practices of digital nomads, who lead location-independent lives and rely heavily on technology to both conduct their work and maintain social relationships~\cite{nash2018digital}. These individuals leverage various technologies to support their flexible work arrangements, including platforms for finding employment and tools for communicating with distributed teams, but encounter challenges such as reliable internet connectivity and access to power~\cite{sutherland2017gig}. Nomadic lifestyles can also complicate digital nomads' efforts to maintain social relationships and engage with local communities, making online communities and co-spaces important for supporting professional and social needs~\cite{lee2019social,liegl2014nomadicity}.

Within this broader research area, researchers have investigated the role of technology in vanlife. Vanlifers are individuals who primarily live in converted cargo vans and camper vans, either full-time or part-time, and maintain a hyper-mobile lifestyle while working from within their vehicles~\cite{gretzel2019vanlife}. While vanlifers' experiences overlap with digital nomads in general, vanlife represents a distinct design space for technology due to the constraints and affordances of vans as a place for life and work. Rizvi et al.~\cite{rizvi2021provocations} found that vanlifers invested significantly in technology and identified four key pillars constituting the design space for technology in vanlife: limited power, variable internet connectivity, limited space, and limited disposable income. Based on these considerations and vanlifers' current technology practices, they proposed provocations for technologies that support van lifers' social relationships within the community and personalization of their living spaces. Complementing this work, design research has explored the material practices of van conversion and customization. Desjardins and Wakkary~\cite{desjardins2016living} conducted an autobiographical design project documenting a camper van conversion, highlighting the importance of do-it-yourself practices and the role of online community resources in the reconfiguration process. Suzuki et al.~\cite{suzuki2023advancing} illustrated how prototyping tools can help vanlifers explore and design van interiors, allowing them to balance functional requirements and personal preferences within the spatial constraints of the vehicle. These studies demonstrate the importance of technology for individuals with mobile lifestyles, both for navigating everyday life and maintaining a sense of community. For vehicle dwellers specifically, these studies point to the important role of technology in configuring vans into livable spaces.

Critically, prior work in this area has mostly included participants who choose vehicle dwelling as a desired lifestyle~\cite{odom2019diversifying,stampf2024deriving}, spend significant time and resources iterating upon and upgrading their vehicles~\cite{zafiroglu2007digital,desjardins2016living}, and feel connected to a broader community of other vehicle dwellers both in-person and online~\cite{rizvi2021provocations,gretzel2019vanlife}. Additionally, existing research focuses on larger vehicles such as vans and RVs, overlooking the spatial and infrastructural constraints of cars that may shape technology adoption and use. Our work addresses these gaps by surfacing car dwellers' technology practices and needs, contributing to an understanding of how technology can better support vehicle dwellers with differing resources, constraints, and orientations to vehicle living.

\subsection{Infrastructure}
Research in HCI has increasingly examined infrastructure and infrastructuring to understand how people navigate social, technological, and physical systems in their daily lives~\cite{lyu2025systematic}. Star and Ruhleder~\cite{star1994steps} conceptualize infrastructure as relational and embedded within other social and technological systems, and highlight that infrastructures largely remain invisible until breakdowns occur. When infrastructures fail, people adapt by appropriating existing systems and creating new infrastructures to meet their needs~\cite{mark2009repairing}. \textcolor{blue}{Vertesi's notion of ``seams'' offers a useful analytical vocabulary for understanding points of breakdown by surfacing the heterogeneity of multi-infrastructural environments~\cite{vertesi2014seamful}. To accomplish local and mundane tasks, people must then ``patch'' multiple infrastructures to bring them into alignment. Using the notion of seams and patches, McClearn et al.~\cite{mcclearn2024security} investigated how people in Lebanon navigate multiple simultaneously failing infrastructures to secure basic needs.} Infrastructures can be challenging to patch~\cite{veeraraghavan2021cat} and require continuous infrastructuring, which refers to the ongoing work of building, maintaining, and repairing infrastructures~\cite{pipek2009infrastructuring}. In periods of prolonged disruption, people engage in ``routine infrastructuring'' to sustain the activities of everyday life~\cite{semaan2019routine}. \textcolor{blue}{The ability to align multiple infrastructures to carry out routine practices constitutes a form of ``infrastructural competence''~\cite{sawyer2019infrastructural,semaan2019routine}.} Collectively, these studies demonstrate the value of infrastructure as an analytical lens for understanding sociotechnical processes across diverse contexts~\cite{robinson2015examining,semaan2012facebooking}, including how marginalized populations creatively assemble multiple infrastructures to meet their needs~\cite{wilcox2023infrastructuring}. 

\textcolor{blue}{Since infrastructure failures typically affect large groups of users, much of the literature on infrastructure focuses on large-scale, community-level efforts~\cite{lyu2025systematic}. However, there have been  calls to examine individual-level efforts to resolve the failures of infrastructure and support the invisible work that individuals perform to make infrastructures work for them~\cite{chen2019unpacking,gui2018navigating}. Gui and Chen~\cite{gui2019making}, for example, uncover individual patients' infrastructuring work to navigate the healthcare infrastructure, drawing attention to how this work is highly localized, individualized, and ephemeral as it meets individual needs without changing the healthcare infrastructure itself. In contrast to prior work that examines how individuals navigate breakdowns within specific infrastructures (e.g., healthcare), or community-level efforts to patch system-wide failures (e.g., ~\cite{paudel2024aftermath,zehrung2025how}), this study highlights the infrastructuring work of individuals to patch multiple heterogeneous infrastructures into alignment to meet their basic needs. Moreover, this work builds on the literature to shed light on how uneven access to resources and broader conditions of sociospatial exclusion shape infrastructuring practices.}

\section{Methods}
To investigate the lived experiences, challenges, and perspectives of car dwellers, we conducted a thematic analysis of posts and comments from a large online community for individuals living in their cars. Car dwellers may be reluctant to disclose their housing status due to the stigmatized and often criminalized nature of vehicle dwelling, making traditional recruitment methods challenging. Online communities therefore provide valuable insight into the real-world experiences of car dwellers, representing a first step toward better understanding the needs of this overlooked population.

\subsection{Data Source}
As a large-scale, publicly accessible platform, Reddit offers rich data on individuals’ diverse lived experiences and perspectives. The platform's pseudonymous nature lends itself particularly well to the study of sensitive topics, as it encourages authentic self-expression and open communication. For example, prior work in HCI and CSCW has demonstrated the value of using Reddit to understand the lived experiences of LGBTQ+ individuals~\cite{modi2025finding,saha2019language}, individuals with disabilities~\cite{motahar2024toward}, and individuals with stigmatized physical and mental health conditions~\cite{jung2025ve,de2014mental,xu2023technology,chopra2021living}. Further, users hold in-depth discussions within the comments under each post, allowing for an understanding of community dynamics as they unfold naturally. 

\subsection{Data Collection}
\subsubsection{Subreddit Selection}
We identified relevant communities by searching for combinations of ``car'' and ``living,'' ``life,'' or ``dwelling.'' This process yielded subreddits such as r/urbancarliving, r/carliving, r/cardwellers, r/priusdwellers, and r/carlife. Among these options, r/urbancarliving demonstrated the highest level of engagement, with 114k members (as of September 2025) and daily posting activity, whereas other subreddits showed minimal active participation. Consequently, we selected r/urbancarliving as our primary data source. 

\subsubsection{Data Extraction and Processing}
We collected posts and comments from the subreddit using Arctic Shift\footnote{\url{https://github.com/ArthurHeitmann/arctic_shift}}, an open-source tool designed for researchers and moderators to access archived Reddit data. While the subreddit was created in 2012, we elected to analyze data from within the last five years at the time of data collection (i.e., January 1, 2020 to January 1, 2025) to examine users' contemporary experiences and practices. This period encompasses significant social, economic, and policy changes affecting vehicle dwellers, enabling us to capture the diverse circumstances and experiences of individuals living in their cars. For instance, the COVID-19 pandemic disrupted housing and employment, which both enabled remote workers to adopt nomadic lifestyles and displaced other people from their homes~\cite{de2021covid,pruss2022long,usich2022vehicular}. Within this timeframe, new policies criminalizing activities associated with homelessness (e.g., sleeping in public spaces) were also implemented, further shaping the experiences of car dwellers~\cite{vehicleresidency2024court}. 

The initial dataset contained 13,785 posts and 347,595 associated comments. After excluding removed, deleted, and bot-generated content (e.g., automoderator posts), the dataset contained 9,635 posts and 339,728 comments. From this corpus, we randomly sampled 500 posts along with their associated 13,405 comments for analysis. Given Reddit's nested comment structure and absence of character limitations, meaningful discussion often occurs in threaded replies. To ensure we understood the context and relationship between replies, we considered each thread (i.e., a post and its associated comments) in its entirety. Since Arctic Shift provided the posts and comments in separate JSONL files, we reconstructed threads using a Python script, saved them as text files, and imported them in MaxQDA\footnote{\url{https://www.maxqda.com/}} for analysis. In the process of analyzing the data, we only included threads that were focused on firsthand experiences and perspectives related to car living. We excluded threads where the discussion was not primarily focused on car living (e.g., general car maintenance), as well as advertisements and survey requests. 

\subsection{Data Analysis}
The first author conducted inductive thematic analysis on the post and comments. Thematic analysis allowed us to develop a rich and nuanced understanding of community members' perspectives and experiences, as well as leave us open to unanticipated insights~\cite{braun2021thematic}. After familiarizing ourselves with the data, we conducted multiple rounds of open and axial coding. We identified codes such as ``rotating parking spots,'' ``cultivating appearances,'' and ``tinting or covering windows'' to capture users' practices and strategies, as well as codes like ``fear of being found out'' and ``lack of support from family and friends'' to capture users’ perspectives and concerns. Then, we grouped the codes into broader themes such as ``remaining undetected as a car dweller.'' Throughout the analysis process, the first author frequently discussed the coding scheme with the last author. If available, we also recorded the user’s geographic location, vehicle type, and any technologies mentioned. After analyzing 125 posts, we started to observe redundant data with few new codes emerging. To ensure we fully captured the themes in the data, we analyzed an additional 25 posts and their associated comments to reach data saturation. In total, we analyzed 150 posts and 2,692 associated comments. 

\subsubsection{Positionality}
The reflexive and interpretive nature of thematic analysis necessitates acknowledgment of researcher subjectivity~\cite{braun2019reflecting}. The research team has extensive experience conducting HCI research in health and wellbeing, with a shared commitment to centering individuals' lived experiences in technology design. The first author contributes expertise in housing insecurity, while the last author contributes expertise in online communities and personal experience living in a recreational vehicle by choice. These backgrounds fostered sensitivity to the complexity of car dwelling, and our goal was to surface diverse perspectives on car living to inform the design of technologies that address car dwellers' varied needs. 

\subsubsection{Ethics}
This study did not require review from the Institutional Review Board, as Reddit data are considered publicly available. However, we carefully considered the ethical implications of analyzing and reporting on users’ sensitive experiences due to the potential for harm~\cite{proferes2021studying}. Following the approach of prior work~\cite{jung2025ve,fiesler2024remember}, we aimed to safeguard users' privacy by carefully rephrasing quotes and quoting users anonymously (e.g., ``one user shared'') to reduce the risk of re-identification.

\subsection{Limitations}
Given the demographics of Reddit users, our findings largely reflect the experiences of car dwellers within the United States. While we did observe geographic diversity across more than 30 states and some participation from Canada and Australia, our findings are limited to perspectives in the Global North. \textcolor{blue}{Local infrastructures and policies shape car dwellers' practices, and future work should investigate the needs of car dwellers across diverse contexts.} \textcolor{blue}{Further, Reddit introduces hidden factors such as age, gender, and race and ethnicity, limiting our understanding of how individuals may experience car living differently based on their identities. For example, individuals with intersectional identities may disproportionately experience surveillance in public spaces.} Our analysis is also limited by the self-selection bias inherent to participation in online communities, as the data only reflect the experiences of individuals who chose to post or comment within a single subreddit. Our findings may not capture the full breadth of car dwellers' technology experiences due to the digital connectivity required for Reddit participation. For instance, we observed \textcolor{blue}{relatively} limited discussion of internet connectivity challenges or mobile data constraints, likely because our sample only included those with sufficient digital access to participate in the subreddit. \textcolor{blue}{Given the limited research available on vehicle dwellers, we view this study as a first step toward understanding the unique needs and challenges of this population. Future work should consider additional methods such as ethnographic approaches to broaden our understanding of this diverse population and how technology might be designed to better support specific communities of car dwellers.}

\section{Findings}
\begin{figure*}[htbp]
    \centering
    \includegraphics[width=0.9\textwidth]{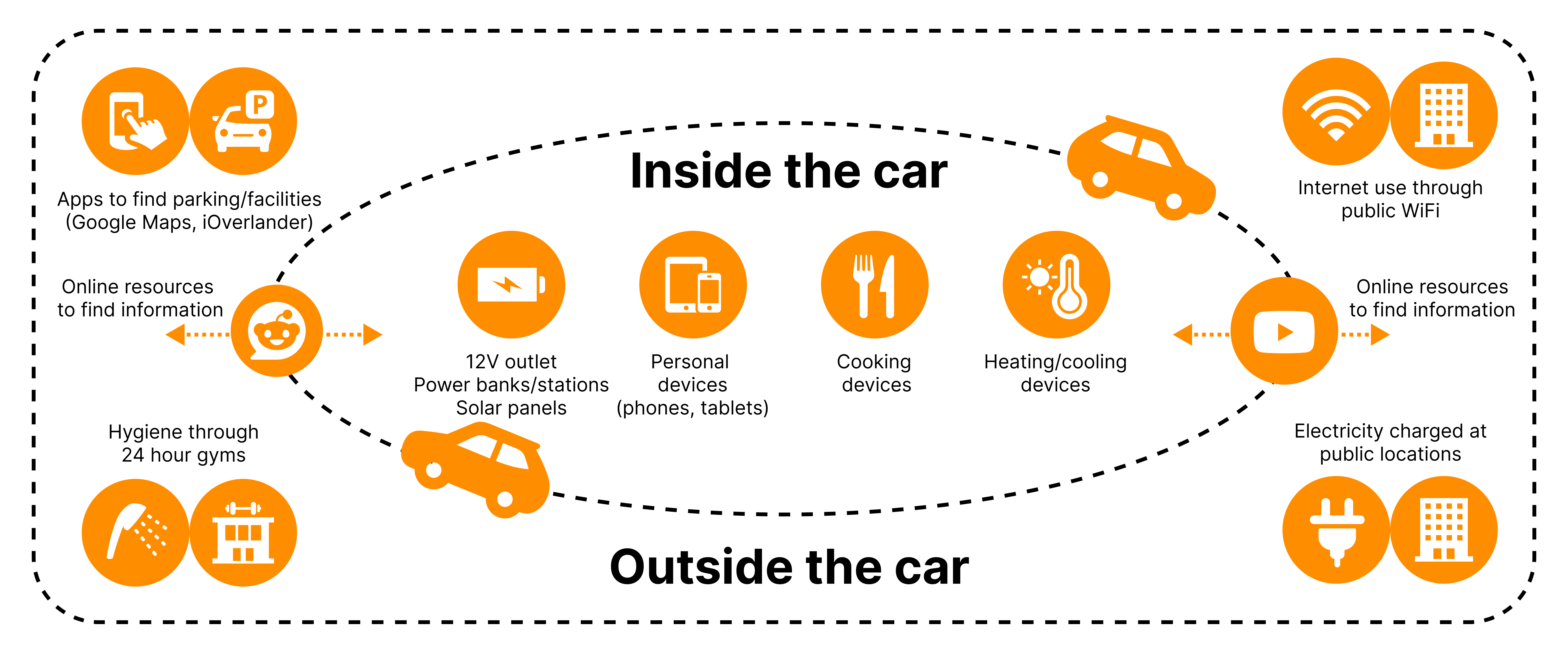}
    \caption{\textcolor{blue}{An overview of the infrastructures inside and outside the car, including the role of different technologies.}} 
    \Description{An oval representing “inside the car” shows icons to represent how car dwellers had access to the car’s 12V outlet, power banks, and solar panels for electricity; personal devices such as phones and tablets; cooking devices; and heating and cooling devices. A rectangle encompassing the oval represents “outside the car,” showing icons to represent the apps car dwellers used to find parking and facilities, their use of gyms for hygiene, public WiFi, and electricity at public locations. Mediating both of these layers are online resources such as Reddit and YouTube.}
    \label{fig:example}
\end{figure*}

Our findings highlight the diversity within the car dwelling community and provide insight into the challenges, practices, and identities that both connected users and contributed to tension among them. \textcolor{blue}{We shed light on the work that users performed to make car living work for them both inside and outside the car, including how they leveraged different technologies to do so (as shown in Figure~\ref{fig:example}).} 

\begin{rachael}
\subsection{Meeting Basic Needs and Making the Car Livable}
\end{rachael}
\textcolor{blue}{Transforming the vehicle into a livable space was challenging and required significant work. A livable space within the vehicle encompassed several key dimensions: the means to perform daily activities such as caring for personal hygiene and cooking, the ability to maintain a safe and comfortable temperature, and an interior that supported sleep and privacy.} To overcome the spatial and infrastructural constraints of the vehicle, users creatively repurposed and modified the car's structure and systems, while drawing upon external infrastructures when their needs exceeded what the car could accommodate. 

\subsubsection{Maintaining Personal Hygiene}
Establishing a sustainable routine \textcolor{blue}{for maintaining personal hygiene} was essential for managing life in the car. \textcolor{blue}{Hygiene impacted both physical health and mental wellbeing, and it reflected users' control over their circumstances (e.g., ``\textit{Once you start letting hygiene slip, it causes a domino effect. Don't ever give that up}).''} \textcolor{blue}{To maintain their personal hygiene, users developed a range of strategies that combined improvised solutions inside their vehicle with reliance on external infrastructure.} Gym memberships were the most frequently recommended option, as they provided relatively low-cost access to showers and toilets at all participating locations. As one user advised, ``\textit{A gym membership was how I showered most days. I also kept two buckets and gave myself a sponge bath on the days that I couldn't shower. With a garbage bag, the bucket also served as an emergency toilet.}'' Users incorporated visits to the gym as part of their everyday routine in order to meet their hygiene needs. However, while gyms were convenient for most cases, users still emphasized the need for backup options when gyms were inaccessible. For example, makeshift bathing systems allowed this user to sustain themselves in between shower opportunities. Likewise, being unable to reach the gym or another bathroom necessitated an emergency waste container to use inside the vehicle. Collectively, these strategies highlight how users relied on a hybrid system wherein they \textcolor{blue}{equipped their cars with limited infrastructure to maintain their hygiene but relied on external facilities when possible.}

\subsubsection{Managing Diet Without a Kitchen}
\textcolor{blue}{Lacking space and access to conventional kitchen facilities, car dwellers faced challenges with food storage and preparation. While dining out was the most convenient workaround, it was financially unsustainable for many users, leading them to develop creative cooking setups within the car. The auxiliary power outlet (commonly known as a cigarette lighter) often served as an entry point for in-car food preparation.} As one user described, ``\textit{I got an electric lunch bag for \$30. It's like a thermal bag with a little hot plate inside. It plugs into my car lighter (12V)... it can even cook raw meat}.'' While electric lunch boxes are typically designed for reheating meals in workplace settings, \textcolor{blue}{users repurposed them as primary cooking devices in the context of car living.} By appropriating the auxiliary power outlet, this user leveraged the car's existing  infrastructure \textcolor{blue}{to support a low-cost, makeshift cooking setup. However, the auxiliary outlet's limited capacity quickly became a bottleneck, as standard 12V outlets in cars operate at low wattage and cannot support high-power appliances such as microwaves. This technical constraint necessitated more extensive infrastructuring work} and higher levels of technical knowledge. One user shared, ``\textit{I picked up a kettle, slow cooker, roaster, and frying pan that are all 12V. Because they are around 200 to 300 watts and my cigarette lighter has a 180w, 15amp capacity, I'll install a dedicated 30amp line to power them.}'' This approach involved recognizing the limitations of the auxiliary power outlet, calculating the electrical requirements of multiple appliances, and expanding the car's infrastructure by installing a separate higher-capacity electrical line to support these appliances. 

Despite the benefits of cooking, users emphasized the safety risks of preparing food inside a vehicle, leading them to supplement with outdoor cooking methods when possible. Concerns included fire safety, condensation and mold, and gas leaks from portable stoves, \textcolor{blue}{which were} compounded by insufficient ventilation and space. To mitigate these risks, some users suggested outdoor cooking methods (e.g., ``\textit{find parks with free grills}''). However, public facilities were not always accessible, particularly in urban areas. As one user shared, cooking at apartment grills seemed dangerous because ``\textit{there's always that one guy who knows when you don’t live there. You're basically playing hide and seek.}'' In these mixed public-private spaces, users were concerned about being perceived as out of place, contributing to a sense of anxiety and avoidance. Overall, users weighed the trade-offs between cost, safety, and discretion when \textcolor{blue}{meeting their basic needs}.


\subsubsection{\textcolor{blue}{Maintaining Safe Temperatures Within the Vehicle}}
\textcolor{blue}{To sleep safely and comfortably in their cars, users needed to be able to cope with extreme temperatures. }As one user bluntly stated, ``\textit{You can die in your car from heatstroke. Winter weather is just as bad. You can freeze to death depending where you are.}'' \textcolor{blue}{Considering these health risks, users developed workarounds to maintain safe temperatures within the vehicle. One approach involved personal temperature regulation: ``\textit{It's easier to heat yourself than it is to heat the whole vehicle. The best way I've found is a power bank, a heating blanket, and a good sleeping bag.}'' This solution relied on a small external power bank to operate a low-wattage heating device, offering a convenient and relatively affordable way to cope with cold weather. To recharge power banks, users relied on a combination of the car's auxiliary power outlet and outside locations such as public libraries, coffee shops, and gyms.} Individuals with greater financial resources invested in more extensive setups that focused on regulating the vehicle's interior temperature. One user described, ``\textit{I made a 1280WH powerstation, but quickly learned that isn't enough in this heat. I'll be upgrading soon so that I can run portable AC at night. I'm thinking of building my own solar panels too.}'' \textcolor{blue}{Although power stations can support energy-intensive appliances such as air conditioning units, high energy consumption necessitated frequent recharging. Adding the power station therefore introduced dependencies and generated further infrastructuring work, as the user needed to expand their system with solar panels to reliably charge the device and sustain overnight climate control.} One persistent concern was that modifications intended to make car dwelling more comfortable could also compromise stealth. As one user explained, ``\textit{I'm concerned about someone seeing a running fan or AC exhaust vent. I also have a bunch of glowing power banks in the car that I worry would be visible. I found some cheap vent covers, but that might look suspicious too.}'' This tension underscored a key challenge of car living: balancing customization and comfort with the need to blend in and avoid detection.

\subsubsection{\textcolor{blue}{Reconfiguring the Vehicle Interior for Sleep}}
\textcolor{blue}{
Along with maintaining safe temperatures, users needed to reconfigure the interior in order to sleep comfortably within the limited space of the vehicle. To learn how to modify the interior, users frequently turned to online resources. As one user explained, ``\textit{I took out the back seats and put in a little bed... look up your exact car's rear seat removal on YouTube, it's easy.}'' Because vehicle makes and models varied, users sought out vehicle-specific guidance on platforms such as YouTube to meet their specific needs. In addition, users sometimes shared photos of their setups on the subreddit to assist others with their customization efforts. However, even with creative interior layouts, some users found it challenging to create a sleeping area while storing all their possessions, leading them to consider external storage units. As one user shared, ``\textit{I use a storage unit for most of my stuff so that my car stays clean. The car should only be used for sleeping and transportation.}'' This case highlights how the spatial limitations of vehicles pushed users to supplement with external infrastructures in order to meet their needs.}

\textcolor{blue}{Users also modified the vehicle interior to cultivate a sense of domesticity and separation from the outside world, which was particularly important for restful sleep. As one user emphasized, ``\textit{you can only sleep fully when you're somewhere safe and familiar. Your car has to become that place.}''  Window covers and tints, for instance, could transform the vehicle into a private space by creating visual barriers between the vehicle's interior and the outside world. One user described the effect: ``\textit{It's such a different world when you can cover your windows and really feel like you have a small apartment inside your car.}'' Personal devices such as phones, tablets, and laptops also played a role in converting the interior to a home-like environment by supporting leisure activities and routines around sleep. For example, one user described their routine of unwinding in their ``\textit{car cave}'' with ``\textit{a tablet hooked over the visor playing a movie or YouTube and laptop for playing games.}'' }

\subsection{\textcolor{blue}{Finding Safe and Strategic Parking}}
\textcolor{blue}{While the car itself served as a private living space, users needed to strategically position their vehicles within public spaces to carry out their everyday routines. Parking was not simply a matter of finding an available space, but rather an ongoing process of ensuring personal safety while maintaining proximity to the infrastructures needed to sustain everyday life. To identify potential parking locations, users relied on various digital technologies including the subreddit, mobile applications, and digital maps.}

\subsubsection{\textcolor{blue}{Defining Safe and Strategic Parking}}
\textcolor{blue}{Users selected parking spots based on two primary considerations: the perceived safety of the location and its strategic proximity to external infrastructures necessary for daily life.} They understood safety as both protection from harm and the ability to inhabit a space without the constant threat of removal. As one user explained: ``\textit{The hardest part about doing this is where to park, and safely. As much as I try to find safe places and be stealthy, there's always a looming feeling of getting the knock, being robbed, or harassment.}'' Here, safety refers not only to protection from crime and harassment, but also to avoiding ``\textit{the knock,}'' which is a term used to describe being identified by police or private security. At best, the knock disrupted users' routines by forcing them to relocate; at worst, it resulted in being towed, cited for violating overnight parking laws, or experiencing harm. \textcolor{blue}{To minimize the risk of such encounters, most users rotated parking spots (e.g., ``\textit{never stay in one place for more than two to three days. You get noticed''}). Even so, the constant anticipation of this disruption  contributed to a pervasive sense of anxiety and uncertainty.} 

\textcolor{blue}{In addition to safety considerations, users strategically selected parking spaces based on their proximity to facilities needed for managing daily life, such as gyms for hygiene, storage units for belongings, and establishments offering WiFi for connectivity. One user shared, ``\textit{I stopped by my storage unit every few days to pack up some work clothes. I'd sleep in the college parking garage, go to the gym, take a shower, and change for work. My office never knew I was living in my car.}'' To make car living work, users needed to align their routines with external infrastructures, which shaped their parking choices. Similarly, another user described their approach to maintaining internet access while parked overnight: ``\textit{I'd visit coffee shops and restaurants in the city, get the WiFi password, and go back after closing hours to use the signal from my car. So wherever you are, you can find an already marked WiFi spot.}'' This strategy involved creating a network of parking locations across the city to stay connected to digital infrastructure. These accounts also suggest that car dwelling did not necessarily imply nomadism. Although users rotated parking spots for safety, many remained within the same geographic area and developed routines anchored to particular facilities.}

\subsubsection{Leveraging Digital Technologies to Find Parking}
To support their search for safe \textcolor{blue}{and strategically located} parking, users relied on the subreddit, mobile applications, and mapping tools. For general guidelines, users frequently sought and shared advice within the subreddit. General strategies entailed parking at public facilities (e.g., rest stops, campsites, and federal lands), businesses that are known to allow overnight parking (e.g., 24-hour gyms and hotels), hospitals, and streets in residential areas without parking enforcement. While some posters sought parking recommendations in specific geographic areas, responses typically offered broad advice rather than exact locations. For more localized guidance, users recommended mobile applications such as iOverlander and RVParky, which rely on user contributions to provide information on parking locations and facilities. As one user advised, ``\textit{Try iOverlander and read the reviews. People mention if it's a great place to stay or if they felt unsafe. The app is for RVers, but I've seen people car camping at many of the places.}'' Although these applications were originally designed to support trip planning for RVers, car dwellers used them to navigate everyday life. In particular, they valued learning from other individuals' personal experiences to help them assess the safety of potential parking spots and make more informed decisions about where to stay. \textcolor{blue}{Another user commented that ``\textit{iOverlander can be useful in finding other resources such as water, shower facilities, and truck stops,}'' suggesting that these applications could also help users plan around access to external infrastructures. }

\textcolor{blue}{While shared resources were beneficial,} users stressed the importance of finding their own unique parking spots. \textcolor{blue}{As one user explained,} ``\textit{most people won't list their best spots [on iOverlander]}.'' \textcolor{blue}{Some users even recommended withholding information about safe and strategic parking locations in order to protect the viability of well-located parking spots.} As one user advised: ``\textit{Be cautious about sharing information about good places to park at night. Sometimes the word gets out and then it fills up too quickly, and whoever has the legal right to the space gets upset and kicks everybody out.}'' Users had to balance their desire to support other community members with the need to safeguard their own spaces and wellbeing, suggesting that localized knowledge could be a valuable resource. To independently locate parking spots, users relied upon more exploratory methods such as leveraging digital maps. \textcolor{blue}{One user shared, ``\textit{I start with Google and Apple Maps. I use the satellite view to scout potential places, then use the street view to check any potential problems and to try and read the parking signs. If it looks good, I'll scout the spot in-person during the day.}}'' Instead of relying on information in the subreddit or crowdsourced reports, this approach required an individual assessment of the built environment. While the satellite view provided a high-level overview of potential parking areas, the street view allowed users to analyze details of the environment before attempting to park there. \textcolor{blue}{Moreover, digital maps allowed users to develop an understanding of where parking spots were located in relation to other facilities.}

\subsection{Navigating Identity and Community}
Users' experiences of car dwelling were not only about managing material conditions, but also about negotiating questions of identity and belonging. The reasons why users entered car living and their ability to manage daily life shaped how they understood themselves, their place within the broader community of car dwellers, and their position in society at large. 

\subsubsection{\textcolor{blue}{Motivations for Car Living}}
Motivations for car living fell into two distinct categories: as a necessity or as a lifestyle choice. \textcolor{blue}{For necessity-driven users, car living emerged as a temporary survival strategy in response to economic precarity and housing instability.} As one user explained: ``\textit{I'm falling behind on bills and rent. I've been looking into my options over past few days, and the most feasible option is turning my car into a place to live.}'' Faced with high housing costs and insufficient income, these individuals turned to their vehicles as an accessible and immediate solution when conventional housing became unattainable. Although car living offered certain advantages over alternatives such as homeless shelters (namely privacy and autonomy), necessity-driven users framed car living in terms of what it lacked \textcolor{blue}{both materially and psychologically}. As one user reflected after two years of car dwelling: ``\textit{I'm worn out. I want a nice bed and a real kitchen to cook in. There's always the `what ifs' like getting sick or injured, robbed or assaulted, or even arrested. It's a lot of stress.}'' \textcolor{blue}{Overall, necessity-driven users resorted to car living as a temporary solution to financial strain and housing issues, with the goal of returning to conventional housing.}

\textcolor{blue}{In contrast, lifestyle-driven users framed car living as an intentional and preferred way of life, motivated by desires for freedom, mobility, and financial flexibility.} One user shared: ``\textit{I've decided to leave my apartment and live in my car by choice. There's this freedom I've been craving, and this gives me a chance to hit pause on traditional living, save money, and have some adventures.}'' Another emphasized the positive impact on their wellbeing: ``\textit{Car living was a choice for me. My mental and physical health have greatly improved. I also have a better quality of life since I get to keep most of my income.}'' For these individuals, car living represented a liberating alternative to the constraints of conventional housing. Other users were motivated by employment-related reasons, such as the ability to pursue better work opportunities. As one user explained, ``\textit{Car living gives you the freedom to put yourself in places that can offer opportunities... I can save money since hourly wages there are higher.}'' Unlike necessity-driven users, lifestyle-driven individuals often had no intention of returning to conventional housing, viewing car dwelling as a desirable long-term arrangement that enhanced their overall quality of life. 

While the categorizations of necessity-driven and lifestyle-driven overlook the precise reasons for car living, we emphasize the main distinction as the element of choice (i.e., whether car living was undertaken out of necessity or desire). In the following sections, we describe how motivations for car living shaped users' identities and perceptions of their living situations.

\subsubsection{Houseless, Not Homeless}
\textcolor{blue}{Lifestyle-driven users often rejected the premise of homelessness, instead perceiving their cars as nontraditional homes. }As one user reflected, ``\textit{I'm doing this by choice and had lots of time to prepare. I don't know what it's like to be homeless. Homeless and houseless are two different things.}'' Here, the key distinction between houseless and homeless derived from the user's sense of agency. Becoming houseless was framed as a choice, whereas becoming homeless was understood as the result of adverse circumstances. Houseless described only the nature of the physical dwelling, without implying the loss of home. Importantly, these users had the time and resources to transform their cars into comfortable living spaces and took pride in their cars as legitimate homes. One user stated, ``\textit{I tell everyone [about my lifestyle]. I'm really proud of my tiny home on wheels.}'' Users who identified as houseless openly disclosed their living arrangements and often rebuffed well-intentioned attempts to classify them as distressed or in need of help. As one user remarked, ``\textit{People try to throw a pity party for me and feel bad for me, but I'm living my best life.}'' By making it clear that they did not require assistance, these users reinforced their autonomy and the legitimacy of their chosen way of living. 

\subsubsection{Homeless, But Not Part of the Homeless Community}
\textcolor{blue}{In comparison, necessity-driven users tended to identify as homeless and viewed car living as a destabilizing experience.} As one user recalled, ``\textit{For me, the first night [in the car] was filled with sadness, feelings of total failure, loneliness, and fear.}'' For such users, the car represented displacement rather than a new home, and the transition was disruptive to their sense of self-worth, stability, and community. As a result, they often felt ashamed of their living situations and concealed their circumstances from their social circles in an effort to avoid judgment. As one user cautioned, ``\textit{Don't tell anyone in real life that you are homeless. The stigma is real.}'' However, acknowledging one's circumstances as homelessness did not necessarily translate into acceptance of this new identity or a sense of community belonging. Rather, to cope with their loss of social status and preserve aspects of their prior social identity, many users distanced themselves from people experiencing homelessness (PEH) in everyday life. As one user explained, ``\textit{My whole life depends on nobody realizing I'm homeless so no asking for help, no holding a sign, no `looking homeless' (shave, shower, and dress normally etc). I avoid other homeless people as much as I can.}'' Although this user recognized themselves as homeless, they did not want to affiliate with other PEH because doing so could jeopardize their personal and professional standing. They believed that accepting services would also expose their status, yet by rejecting assistance and avoiding others in similar situations, they excluded the possibility of offline support. Another user similarly distanced themselves from PEH: ``\textit{The other homeless really ruined things for me. I don't steal, use drugs, beg, bother anyone, or damage property, but the others did, and I struggled because of it. Then my state passed a law that made sleeping overnight a felony.}''  While such users acknowledged that they were homeless, they believed that they were different from ``other'' PEH because of their ability to maintain an image of normalcy and manage aspects of daily life (e.g., working and maintaining personal hygiene). In attempting to differentiate themselves, users inadvertently reinforced prevailing stereotypes of homelessness such as uncleanliness, deviance, and disorderly conduct, reproducing the same stigmatization that they themselves feared. Importantly, many users viewed their situations as temporary and encouraged each other to ``\textit{ignore the homeless mindset,}'' as one user described, suggesting that they viewed themselves as a distinct community. 

\subsubsection{In-Between Identities}
At the same time, the distinction between homeless and houseless was not always straightforward, as users' identities could shift in response to their changing circumstances and lived experiences. That is, users' initial motivations for car living were not necessarily reflective of their current identity and self-perception. Some lifestyle-driven users who initially embraced car living began to identify as homeless when it was no longer a matter of choice. As one user shared, ``\textit{I started by choice right before covid. I think it has slipped into homelessness as I'm not earning enough income to leave this life now that I want to.}'' In this case, the user's perceived loss of agency marked the shift from houseless to homeless. Conversely, some necessity-driven users who initially identified as homeless found that the process of adapting their cars into livable spaces reshaped their self-perception. As one user explained, ``\textit{I'm getting used to car life and looking at the bright side. I'm finding that day by day it gets easier. I'm really enjoying not having to pay rent and I'm also loving the freedom I have.}'' For these individuals, improving their ability to manage daily life and reconfiguring their cars as homes could be an empowering process, shifting their view of themselves and car dwelling. Taken together, these accounts highlight the blurred boundaries between homelessness and houselessness, suggesting that identity was shaped less by users' physical dwellings or their initial motivations for car living, and more by their sense of agency and \textcolor{blue}{ability to manage daily life.} 

\subsubsection{Misalignments Within the Car Dwelling Community}
While the subreddit served as a valuable place to exchange strategies and peer support, there was sometimes a misalignment between users due to identity-based differences. In particular, necessity-driven users perceived inadequate support from lifestyle-driven users, whom they viewed as minimizing the realities of car living as a form of homelessness. One user reflected, ``\textit{When I had to live in my car, I came to this sub. While most of the posts were  helpful, a lot of posts `glamorize' this. To me personally, I wasn't living a cool lifestyle. I was fucking homeless.}'' When lifestyle-driven users boasted about their customized setups and touted the benefits of car living, users who identified as homeless felt alienated, as they were struggling to survive and secure stable housing again. The advice shared within the subreddit by lifestyle-driven users was often not helpful for them, as it was misaligned with their financial resources and priorities. Another user articulated this divide within the community: ``\textit{Many of the super positive people on here have an easy way out of the lifestyle, but most of us are stuck doing it while we try to save enough to get out of our cars. I would do anything to have a home and privacy again.}'' The idealistic outlook of lifestyle-driven users came across as insensitive to those who felt trapped in car living. For example, some lifestyle-driven users encouraged necessity-driven users to simply change their perspective: ``\textit{With the right mindset, you can survive anything. Just embrace it as an experience of growth.}'' At the other end of the spectrum, \textcolor{blue}{lifestyle-driven users} often could not understand the frustration of necessity-driven users. As one user shared, ``\textit{I don't know why people say car living is hard. I've been doing it and it's not bad at all. Get a storage unit, a mailbox, cooler for food, wireless fans, gym membership etc.}'' While these amenities can certainly make car living more tolerable or even enjoyable, they were often out of reach for individuals with limited financial resources and misaligned with their immediate survival needs and long-term housing goals. Overall, necessity- and lifestyle-driven users had fundamentally different perspectives on car living, which resulted in the needs of necessity-driven users being overlooked and under-supported within the subreddit.

\section{Discussion}

\subsection{Car Living as a Life In Between}
Our findings suggest that car living can be best understood as a life in between. Car dwellers occupy a liminal state across multiple dimensions: socially, between home and homelessness; spatially, between private and public spaces; and temporally, between short- and long-term arrangements without clear endpoints. Car living is therefore not a static condition, but an ongoing process of adaptation as car dwellers navigate shifting circumstances and identities. 

Prior work has approached vehicle dwelling from two dominant perspectives. Social science research has primarily examined vehicle dwelling through the lens of homelessness, positioning vehicles as temporary, precarious, and inadequate forms of shelter~\cite{giamarino2024lives,giamarino2023planning,calhoun2023safety}. In contrast, HCI research has explored vehicles as alternative domestic spaces that expand the concept of home itself~\cite{desjardins2016living,odom2019diversifying}. These divergent framings have shaped assumptions about the role of technology for vehicle dwelling. If vehicle dwelling is understood as a form of homelessness, then technology might aim to help people cope with the challenges of homelessness and ultimately exit to stable housing~\cite{woelfer2010homeless}. If vehicle dwellings are understood as alternate sites of domesticity, then technology might instead aim to support personalization, social routines, and improving life within the vehicle~\cite{odom2019diversifying,suzuki2023advancing}. 

Our findings suggest that these approaches are not mutually exclusive, as car dwellers often navigate both of these realities at once and hold nuanced views on their living situations. Necessity-driven car dwellers, for example, may regard car living as a form of homelessness while simultaneously rejecting homeless identities and social services~\cite{preece2020living}. Despite the involuntary nature of their living situation, they devote considerable effort to personalizing and transforming their vehicles into livable spaces. Conversely, lifestyle-driven car dwellers may frame their cars as nontraditional homes yet still acknowledge elements of necessity shaping their decisions (e.g., seeking employment opportunities). \textcolor{blue}{Regardless of initial motivations, all car dwellers must develop }strategies to overcome common challenges such as hygiene and cooking \textcolor{blue}{in light of infrastructrual constraints.} \textcolor{blue}{This ability to bridge infrastructural gaps to carry out routine activities can be conceptualized as a form of ``infrastructural competence''~\cite{sawyer2019infrastructural}.}

\textcolor{blue}{While Sawyer et al.~\cite{pipek2009infrastructuring} conceive of infrastructural competence as a characteristic of professional identity, our findings suggest that infrastructural competence can itself shape an individual’s identity and self-perception. That is, we found that} car dwellers’ identities \textcolor{blue}{could shift according to changes in their circumstances and ability to manage everyday life. }What begins as a lifestyle choice can transform into homelessness if the dweller perceives a loss of agency over their living situation \textcolor{blue}{and struggles to meet their basic needs.} Alternatively, \textcolor{blue}{users' identities could shift from ``homeless'' to ``houseless'' as they developed competence in managing everyday life.} These experiences illustrate the nature of car living as a life in between, where notions of home, identity, and belonging are in flux, \textcolor{blue}{and deepens our understanding of the relationship between infrastructural competence and identity.}

\textcolor{blue}{Importantly, we do not suggest that homelessness is simply a matter of perspective. Rather, we highlight the potential of supporting individuals' infrastructural competence to equip them with the knowledge and skills to better meet their everyday needs and to support their resilience in navigating challenging circumstances~\cite{semaan2019routine}. We also emphasize that opportunities to develop infrastructural competence are uneven. We observed this divide between necessity- and lifestyle-driven users who came to car living with different resources and constraints, which shaped their ability to manage everyday life. Building on HCI and CSCW scholarship that has drawn attention to issues of power and access in infrastructure~\cite{singh2021seeing,lyu2025systematic}, we further extend this work to the context of mobility and dwelling.}

\textcolor{blue}{\subsection{Mobile Dwellings: Infrastructuring a Private Life in Public Spaces}}

\textcolor{blue}{In considering mobility and dwelling, our findings point to the importance of making visible the gaps between private and public infrastructures, particularly in contexts where access to infrastructure is uneven. The ``home'' is typically conceptualized as a private and protected domestic space~\cite{yao2019defending}, yet mobile dwellings complicate this understanding by necessitating engagement with public spaces. Like digital nomads and van lifers, car dwellers must navigate multiple infrastructures and public spaces as part of everyday life~\cite{oogjes2018designing,nash2021nomadic}. However, a key difference is that many car dwellers do not have a choice in engaging in this type of infrastructuring work. Digital nomads, for example, can seek out hotels and short-term leases (thereby creating private spaces~\cite{lee2019social}) while van lifers can modify their vehicles such that they require limited engagement with public infrastructures (e.g., equipping their vans with a bathroom~\cite{rizvi2021provocations}). In contrast, car dwellers must contend with what scholars in housing studies refer to as the paradox of ``attempting to live private lives in public spaces''~\cite{robillard2023100}, introducing a different set of challenges than those faced by other mobile populations previously studied in HCI. 
}

\textcolor{blue}{While digital nomads and van lifers are mobile by choice, many car dwellers are instead ``fixed in mobility''~\cite{jackson2012fixed}. That is, their mobility is neither free nor random, as it is shaped by access to essential infrastructures as well as forms of surveillance~\cite{jackson2012fixed}. People identified as homeless, for example, are subject to different rules and exclusionary practices than mobile knowledge workers when they try to access digital infrastructures for connectivity~\cite{humphry2021looking}. These dynamics are further complicated when we consider that some users appeared to be geographically constrained (e.g., due to their employment status) and remained within the same city or locality. In terms of information-sharing, for example, nomadic workers often share knowledge of local infrastructures to support other nomads~\cite{mark2010making,lee2019social}. Because they move frequently, sharing this information does not compromise their own access or work practices. In contrast, we found that car dwellers who are geographically constrained view knowledge of local infrastructures as valuable information that, when shared, can be detrimental to their own practices. For instance, safe and strategically located parking spots within a city are limited, and sharing information widely about a spot can attract too much use and lead to increased surveillance by law enforcement. This finding resonates with prior work indicating that people experiencing homelessness only share information with each other when they are not competing for the same limited resources~\cite{hersberger2001everyday}. By surfacing this tension, we highlight the need to consider different types of supports and technologies to support people who are not necessarily mobile by choice, in contrast to much of the literature in HCI and CSCW. To identify opportunities for technology, we use an infrastructuring lens to unpack the work car dwellers performed to sustain everyday life and the role of technology in this process.}

\subsection{Unpacking the Infrastructuring Work of Car Dwellers}
\textcolor{blue}{Car living disrupted users' access to the infrastructures that enabled everyday life in conventional housing (e.g., power grids and sanitation networks). To meet their basic needs in light of this ongoing disruption, car dwellers engaged in ``routine infrastructuring''~\cite{semaan2019routine} work that required them to navigate infrastructures both inside and outside the vehicle. Within the private space of the vehicle, users repurposed the physical and electrical infrastructures of the car to support domestic activities. Outside the vehicle, they leveraged public and semi-public infrastructures such as sanitation infrastructures (e.g., gyms), the electrical grid (e.g., public outlets), and physical infrastructures (e.g., storage units) to complement their practices inside the car. More broadly, they navigated institutional infrastructures such as local parking regulations and digital infrastructures such as WiFi locations, which shaped how they could perform daily activities. In the process of navigating everyday life, users frequently encountered ``seams'' across these heterogeneous sociotechnical infrastructures and creatively patched them to accomplish mundane daily tasks~\cite{vertesi2014seamful}. For example, the 12V direct current output of a car battery is incompatible with cooking devices designed to draw from standard power outlets, leading to workarounds like wiring directly to the car battery. Such adaptations make visible the infrastructural misalignments car dwellers experienced, as well as the invisible work they performed to align them.}

\textcolor{blue}{Further, we found that infrastructuring work was highly individualized and localized because users differed in their geographic locations, vehicle types, and access to resources. Similar to how patients must develop their own knowledge and skills to navigate complex and localized healthcare infrastructures~\cite{gui2018navigating,gui2019making}, users had to undertake individualized and localized endeavors to make car living work for them. However, in contrast  to prior work that examines how individuals respond to breakdowns within specific infrastructures, our study points to a distinct form of infrastructuring work. That is, we draw attention to the infrastructuring work that individuals must perform to align multiple infrastructures that function in isolation but overlap in such a way that they fail to support the practices of everyday life. The infrastructures that car dwellers rely on do not experience system-wide failures (e.g., as in the case of disaster work~\cite{semaan2011technology}). The car still operates and supports mobility; the electrical grid still functions reliably; and parking regulations are enforced as intended by local governments. For individuals living in conventional housing, these infrastructures easily support their daily routines. Yet, for car dwellers, these infrastructures were never intended to align in such a way that supports their needs and routines. These persistent infrastructural seams could only be patched through individual infrastructuring work. Moreover, car dwellers' infrastructuring practices were further complicated by issues of uneven access to resources and public spaces. Car dwellers identified as homeless faced penalization when performing ``life-sustaining acts'' such as sleeping in their vehicles in public spaces, increasing the burden and the stakes of their infrastructuring work~\cite{aykanian2019criminalization}. We encourage future work to examine how marginalized individuals navigate multiple, layered infrastructures that were never intended to support them, and in some cases, were designed to exclude them.}

\textcolor{blue}{Lastly, we highlight the critical role that digital technologies played in both supporting and generating car dwellers' infrastructuring work. On the one hand, digital technologies helped users access knowledge and learn skills needed to make car living work for them in their localized context. Video tutorials, for instance, supported users in learning how to reconfigure the specific vehicles they owned to support everyday activities, while mobile applications helped them navigate spaces outside the car to find safe and strategically located parking. Likewise, personal devices contributed to transforming the inside of the car into a private and domestic space by supporting leisure activities. On the other hand, using these technologies also revealed additional seams and generated further infrastructuring work. Accessing online resources required internet connectivity, which in turn shaped some users' parking practices as they needed to maintain connectivity from the car. These practices reveal how car dwellers' everyday lives were structured around managing the boundaries between private life inside the car and public infrastructures outside the car, as well as the role of technology in mediating these boundaries.}

\subsection{Implications for Technology Design}
\textcolor{blue}{In designing for car dwellers, particularly those who are necessity-driven or identify as experiencing homelessness, we draw upon Kaziunas et al.’s concept of ``precarious interventions'' to consider the risks of introducing interventions for marginalized communities~\cite{kaziunas2019precarious}. Considering the criminalization of homelessness in public spaces~\cite{o2012varieties,aykanian2019criminalization}, as well as many users' desires to conceal their status as homeless, we caution against assuming that increased visibility or institutional engagement is inherently beneficial. Instead, we identify opportunities for technologies to support car dwellers' individual infrastructuring practices and infrastructural competence, helping reduce the burdens they face in making everyday life work.}

\subsubsection{\textcolor{blue}{Platforms for Knowledge Sharing}}

\textcolor{blue}{Our findings revealed that users relied on a combination of online resources to learn how to reconfigure their vehicles and perform everyday activities, yet they still faced challenges in translating this knowledge to their local and individual context, pointing to the need for more tailored support. Drawing inspiration from the accessibility literature on building do-it-yourself solutions~\cite{cha2025dilemma}, we suggest that collective platforms could facilitate the exchange of practical knowledge about vehicle modifications. Information on transforming cars into livable spaces is currently distributed across different online spaces (e.g., YouTube, subreddits, and Facebook groups), making it challenging for car dwellers to find information suited to their particular needs. A shared platform could enable users to share detailed guides, photos, and videos of their setups, while organizing them with relevant metadata such as vehicle type, budget level, and primary concerns (e.g., maintaining privacy).} Given the variability in automotive safety standards and regulations across countries, such a platform could also incorporate location-specific guidance to ensure that suggested modifications comply with local laws.

\subsubsection{\textcolor{blue}{Tools for Personalization and Planning}}

\textcolor{blue}{We also see opportunities for mobile applications to support car dwellers in envisioning and their interiors while making visible the seams that they are likely to encounter. Although prior work has demonstrated how prototyping tools can support vehicle design~\cite{suzuki2023advancing}, these approaches primarily address issues of spatial layout and do not fully account for the infrastructural requirements of meeting basic needs. For instance, while there might be physical space for an electric stove, it may be unusable without adding a power station as well. A prototyping tool could surface these dependencies early in the planning process and recommend appropriate modifications.}

\subsubsection{\textcolor{blue}{Supporting the Navigation of Public Spaces}} 

\textcolor{blue}{In addition to supporting car dwellers in managing the private space of the vehicle, technologies could support their navigation of public spaces, helping them align the multiple infrastructures needed for everyday life. While mainstream mapping applications typically prioritize navigation, prior research highlights the value of digital maps that allow users explore their environments and locate spaces for personally meaningful activities~\cite{schade2023mapuncover,dearman2011opportunities}. Car dwellers appropriated digital maps creatively, combining satellite and street views to assess the viability of potential parking locations. Digital maps could further augment these practices by recommending parking locations that align with users' preferences and daily routines. For example, including information about available WiFi, restrooms, and local parking regulations could help car dwellers make informed decisions about where to conduct their daily activities across public and semi-public spaces. }

\subsubsection{\textcolor{blue}{Designing More Supportive Online Communities}}

\textcolor{blue}{Lastly, we suggest considerations for designing online support spaces that better address car dwellers' diverse identities and needs.}  While a straightforward approach might be to create separate communities for necessity- and lifestyle-driven car dwellers, such a division limits opportunities for mutual learning and support. Instead, we draw on insights from HCI research on online health communities, which highlight the value of peer matching as a way to help individuals find relationships that fulfill their emotional and informational needs~\cite{fang2022matching}. Prior work suggests that individuals prefer matches based not on broad labels (e.g., diagnoses), but on fine-grained characteristics such as shared feelings, needs, and attitudes towards their situation~\cite{o2017design}. Building on this perspective, we caution against requiring car dwellers to label themselves (e.g., as homeless) as a method of finding peers, as this labeling could contribute to further stigmatization~\cite{phelan1997stigma}. Instead, online communities might facilitate supportive relationships among car dwellers by matching users according to their goals, resource constraints, orientations towards car living, and temporal outlooks (e.g., whether they prioritize short-term survival or long-term lifestyle optimization). Beyond peer matching, emerging technologies such as conversational agents leveraging large language models (LLMs) show promise for providing personalized support that mirrors aspects of peer interaction~\cite{liu2024compeer}, serving as a valuable addition to peer communities.

\section{Conclusion}
This study examined the experiences of car dwellers through an analysis of an online community, revealing how individuals manage daily life in cars under conditions shaped by necessity or choice. We uncover the infrastructuring work of car dwellers to sustain daily life, identifying how they align multiple infrastructures to meet their needs while simultaneously attempting to conceal their housing status. By surfacing the shared challenges of car dwellers alongside the tensions that emerge from different orientations to car living, we shed light on users' complex identity negotiations and how these dynamics unfold within the community. We then discuss how \textcolor{blue}{this work deepens our understanding of mobility and dwelling, as well as how conditions of exclusion shape individuals' infrastructuring work.} Lastly, we identify opportunities for technologies to support car dwellers in developing infrastructural competence. As vehicle dwelling continues to grow both as a chosen lifestyle and a response to housing insecurity, supporting vehicle dwellers requires nuanced approaches that recognize the diversity of individuals' circumstances, motivations, and identities.

\begin{acks}
We are grateful to the anonymous reviewers for their insightful feedback. This work was partially supported by grants from the UC Irvine Donald Bren School of Information and Computer Sciences Academic Senate Council on Research, Computing, and Libraries (CORCL) and the Graduate Assistance in Areas of National Need (GAANN) Program. 
\end{acks}

\bibliographystyle{ACM-Reference-Format}
\bibliography{main}

\end{document}